\pdfoutput=1
\documentclass[iop,apjl,numberedappendix,twocolappendix,revtex4]{emulateapj}
\usepackage{hyperref}
\usepackage{amsfonts,amsmath,amssymb}
\usepackage{txfonts}
\usepackage{graphicx}
\usepackage{bm}
\usepackage{ctable}
\usepackage{url}
\usepackage{breakurl}
\usepackage{fixltx2e} %% forces figure order to remain fixed (otherwise figure*'s can move out-of-order)
\usepackage[latin9]{inputenc}
\newcommand\nablaex{\tilde{\nabla}}
\newcommand\sbfrac{f_{\hat{b}_{x}>0}}
\newcommand\eperp{E_{\perp0}}
\newcommand\B{|\bm{B}|}
\newcommand\sbgrad{P(|\delta_{\ell}\hat{b}_{x}|>1)}
\newcommand\itemspace{\vspace{0.1cm}\\}
\newcommand\vA{v_{\rm{A}}}
\newcommand\amp{\delta B_{\perp}/B_{0}}
\newcommand\ffold{}

\begin{document}

\title{In-situ switchback formation in the expanding solar wind}

\author{J.~Squire}
\email{jonathan.squire@otago.ac.nz}
\affiliation{Physics Department, University of Otago, Dunedin 9010, New Zealand}
\author{B.~D.~G.~Chandran}
\affiliation{Department of Physics and Astronomy, University of New Hampshire, Durham,
New Hampshire 03824, USA}
\author{R.~Meyrand}
\affiliation{Physics Department, University of Otago, Dunedin 9010, New Zealand}

\begin{abstract}
Recent near-sun solar-wind observations from Parker Solar Probe have found a highly dynamic magnetic environment, permeated by abrupt radial-field reversals, or ``switchbacks.'' We show that many features of the observed turbulence are reproduced by a spectrum of Alfv\'enic fluctuations advected by a radially expanding flow. Starting from simple superpositions of low-amplitude outward-propagating waves, our expanding-box compressible MHD simulations naturally develop switchbacks because (i) the normalized amplitude of waves grows due to expansion  and (ii)  fluctuations evolve towards spherical polarization (i.e., nearly constant field strength). These results suggest that switchbacks form in-situ in the expanding solar wind and  are not indicative of impulsive processes in the chromosphere or corona.
\end{abstract}

\maketitle

%%%%%%%%%%%%%%%%%%%%%%%%%%%%%%
\section{Introduction}

The recent perihelion passes of Parker Solar Probe  (PSP) have  revealed a highly 
dynamic near-sun solar wind \citep{Bale2019,Kasper2019}. A particularly extreme
feature compared to solar-wind plasma at  greater distances 
is the abundance of  ``switchbacks'':  sudden reversals of the radial magnetic field associated with sharp 
%changes 
increases in 
the radial plasma flow \citep{Neugebauer2013,Horbury2018,Horbury2020}. Such structures generally maintain a 
%remarkably 
nearly constant field 
strength $\B$, despite large changes to $\bm{B}$. It remains unclear how 
%they develop 
switchbacks originate and whether they are caused by sudden or impulsive events in the chromosphere or corona \citep[e.g.,][]{Roberts2018,Tenerani2020}.
% Uritsky2017

In this letter, our goal is to illustrate that turbulence with strong similarities to that 
observed by PSP develops from simple, random initial conditions within the magnetohydrodynamic (MHD) model. 
 Using numerical simulations, we show that constant-$\B$ radial-field reversals arise naturally when Alfv\'enic fluctuations 
grow to amplitudes that are comparable to the mean field. 
We hypothesize that the effect is driven by magnetic-pressure forces, which, by forcing $\B$ to be nearly constant across the domain  \citep{Cohen1974,Vasquez1998},
 cause large-$\amp$ fluctuations to become discontinuous \citep{Roberts2012,Valentini2019}.
The effect is most pronounced around  $\beta\sim 1$ (where $\beta$ is the ratio of thermal to magnetic pressure).
Due to the radial expansion of the
solar wind, initially small-amplitude waves propagating outwards from the chromosphere 
naturally evolve into such conditions \citep{vanBallegooijen2011,Perez2013,Montagud-Camps2018}. We solve the MHD ``expanding box'' equations \citep{Grappin1993},
which approximate a small patch  of wind moving outwards in the inner heliosphere. 
We deliberately keep the simulations simple, initializing 
with random superpositions of outwards propagating waves, 
assuming a constant wind velocity, 
%neglecting radial-variation-induced wave reflection effects \citep{Heinemann1980,Velli1993}, 
and using an isothermal equation of state. 
That this generic setup reproduces many features of the turbulence
and field statistics seen by PSP strongly suggests 
that switchbacks form in-situ in the solar wind and are not remnants of impulsive or bursty events at the Sun.

\section{Methods}\label{sec:methods}

%%%%%%%%%%%%%%%%%%%%%%%%%%%%%%%%%%
\begin{figure}
%\begin{center}
%\includegraphics[width=0.98\columnwidth]{\ffold fig_tracks_a}\\\includegraphics[width=0.885\columnwidth]{\ffold fig_tracks_b}\\\includegraphics[width=\columnwidth]{\ffold fig_snoopyt}
\includegraphics[width=\columnwidth]{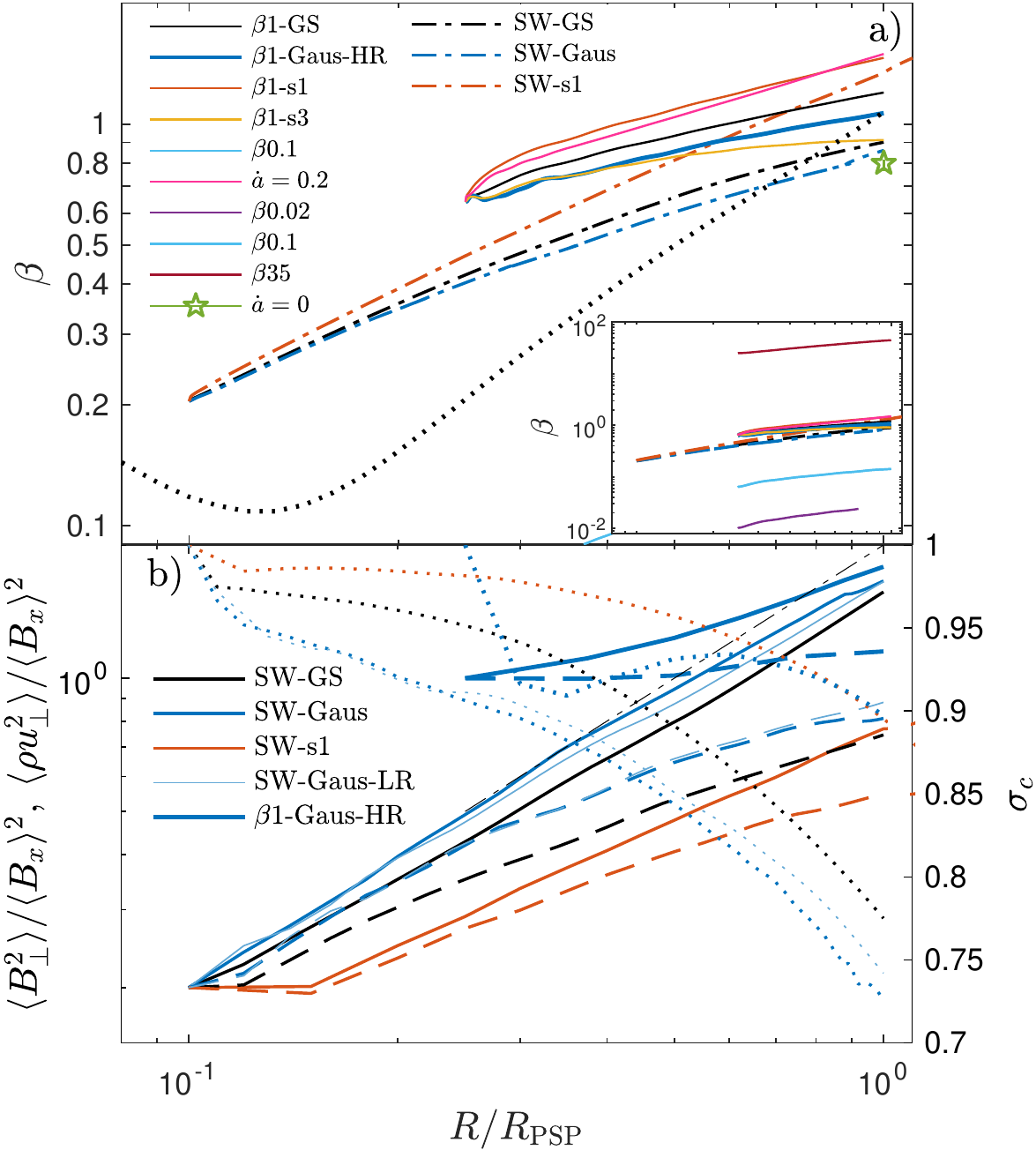}
\caption{Radial evolution of simulation plasma properties. Panel a shows $\beta$
for each simulation as labeled. The Snoopy simulations (dot-dashed lines) crudely approximate  $\beta$ evolution from the model of \citet{Chandran2011}  adapted to match the slow wind seen during PSP's first perihelion \citep[see][]{Chen2020}, which is shown with the
thick dotted line. The Athena++ simulations retain nearly constant $\beta$ as they evolve (the inset zooms out to show $\beta0.02$-GS, $\beta0.1$-GS, and  $\beta35$-GS).
Panel b illustrates the evolution of the normalized  fluctuation amplitude, $\langle B_{\perp}^{2}\rangle/\langle B_{x}\rangle^{2}$ (solid lines) 
and $\langle\rho u_{\perp}^{2}\rangle/\langle B_{x}\rangle^{2}$ (dashed lines). 
The dot-dashed line (almost directly behind the solid line for SW-Gaus) shows the WKB expectation for waves without dissipation: $B_{\perp}/B_{x}=u_{\perp}/\vA\propto a^{1/2}$.
The dotted lines, which are read with the right-hand axis, show the normalized cross-helicity $\sigma_{c}=\langle\!\sqrt{\rho}\bm{u}\cdot\bm{B}\rangle/\langle\rho u^{2}+B^{2}\rangle$. }
\label{fig:PSPtracks}
%\end{center}
\end{figure}
%%%%%%%%%%%%%%%%%%%%%%%%%%%%%%%%%%

%%%%%%%%%%%%%%%%%%%%%%%%%%%%%%%%%%
\begin{figure*}
\begin{center}
\includegraphics[width=1.0\textwidth]{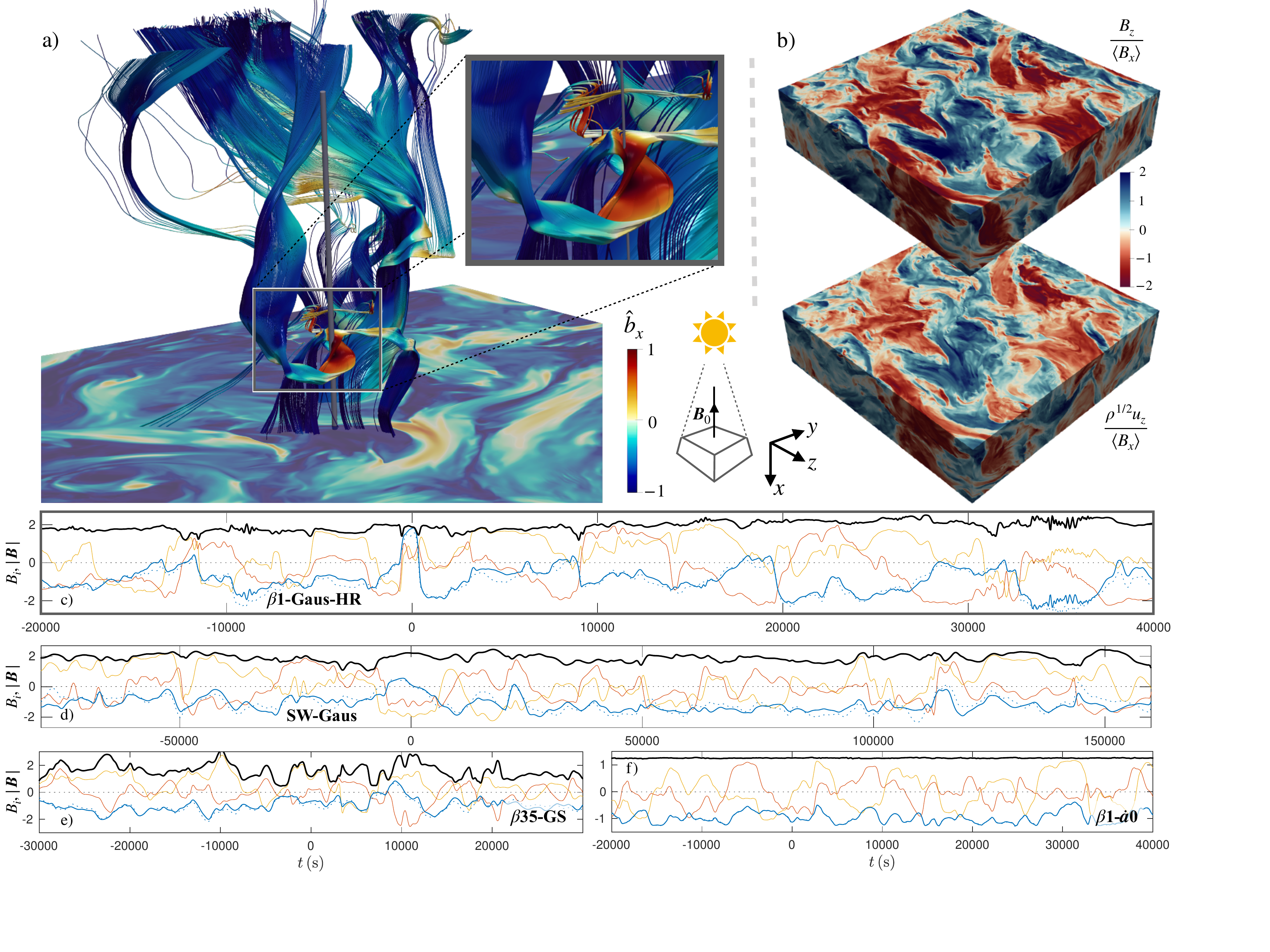}
\caption{The magnetic-field structure in final snapshots, concentrating on $\beta1$-Gaus-HR (panels a--c). Panel 
a illustrates the field-line structure of a strong switchback that appears in the simulation, with the color indicating $\hat{b}_{x}$. 
This switchback appears to be a tangential discontinuity with clear 3D structure to the field lines, 
while a number of rotational discontinuities are  visible in other
regions as sharp bends in the field lines. Right-hand panels illustrate the perpendicular turbulence structure (normalized $B_{z}$ and $u_{z}$),  showing its sharp discontinuities and Alfv\'enic correlation.
%, the correlation between flow and magnetic field, {\bf and the near-homogeneity of the outer-scale motions despite the box having expanded by a factor of 4 over the course of the simulation. }
Panels c--f show simulated ``flybys'' through various simulations, scaled to solar-wind timescales using equations~\eqref{eq:tmap} and~\eqref{eq:f0}.
The flybys sample the direction 
$(1,1/\sqrt{32},-1/2\pi)$ (grey line in panel a) to approximate slow azimuthal motion through a fast radial-wind outflow. Black lines show $\B$, and blue, yellow, and red 
show $B_{x}$, $B_{y}$, and $B_{z}$ respectively, each normalized to $\langle B_{x}\rangle$.  The dotted-blue line shows $\langle B_{x}\rangle+u_{x}\langle 4\pi\rho\rangle^{1/2}$
(normalized by $\langle B_{x}\rangle$), illustrating 
the radial jets  associated with switchbacks \citep{Kasper2019}.}\label{fig:vis}
\end{center}
\end{figure*}
%%%%%%%%%%%%%%%%%%%%%%%%%%%%%%%%%%

We solve the isothermal  MHD equations  in the ``expanding box''
frame \citep{Grappin1993}, which moves outwards in the radial ($x$) direction at the mean solar-wind velocity, while
expanding in the perpendicular ($y$ and $z$) directions due to the spherical geometry. We impose a mean anti-radial (sunward) field $\bm{B}_{0}=-B_{x0}\hat{\bm{x}}$,  with initial Alfv\'en speed $\vA=B_{x0}/\!\sqrt{4\pi\rho}$.
The mass density $\rho$, flow velocity $\bm{u}$, and magnetic field $\bm{B}$ evolve according to
\begin{gather}
\partial_{t}\rho + \nablaex\cdot(\rho \bm{u})=-2\frac{\dot{a}}{a}\rho\label{eq:mhd.rho}\\
\partial_{t}\bm{u} + \bm{u}\cdot\nablaex\bm{u} =  -\frac{1}{\rho}\nablaex\left[ c_{s}^{2}(t)\rho + \frac{B^{2}}{8\pi} \right] +\frac{\bm{B}\cdot\nablaex \bm{B}}{4\pi\rho} - \frac{\dot{a}}{a}\mathbb{T}\cdot\bm{u} \label{eq:mhd.u}\\
\partial_{t}\bm{B} + \bm{u}\cdot\nablaex\bm{B} = \bm{B}\cdot\nablaex\bm{u}-\bm{B}\nablaex\cdot\bm{u} - \frac{\dot{a}}{a}\mathbb{L}\cdot\bm{B}\label{eq:mhd.b}.
\end{gather}
Here $\mathbb{T}=(0,1,1)$ and $\mathbb{L}=(2,1,1)$, $a(t)=1+\dot{a}t$ is the current perpendicular expansion with $\dot{a}/a$ the  expansion rate,  
and $\nablaex \equiv (\partial_{x},a^{-1}\partial_{y},a^{-1}\partial_{z})$ is the gradient operator in the frame co-moving with the expanding flow.  The equation of state is isothermal, but with a sound speed $c_{s}(t)$ that changes in time to mimic the cooling that occurs as the plasma expands
(for adiabatic expansion, $c_{s}^{2}\propto a(t)^{-4/3}$). A detailed derivation 
of equations \eqref{eq:mhd.rho}--\eqref{eq:mhd.b} is given in, e.g., \citet{Grappin1993,Dong2014}. 
The anisotropic expansion causes plasma motions  and magnetic fields to decay in time. Those that vary slowly compared to $\dot{a}/a$ scale as $B_{x}/\!\sqrt{\rho}\propto a^{-1}$, $B_{\perp}/\!\sqrt{\rho}\propto a^{0}$, 
$u_{x}\propto a^{0}$, $u_{\perp}\propto a^{-1}$, with $ \rho\propto a^{-2}$  \citep{Grappin1996}. In contrast, for small-amplitude Alfv\'en waves with frequencies $\gg \dot{a}/a$, $B_{\perp}/B_{x}\propto a^{1/2}$ and $u_{\perp}/\vA\propto a^{1/2}$. 
By appropriate parameter choices, we ensure that it is primarily this latter WKB effect that leads to  large $\amp$  waves  in our simulations. The former non-WKB  effect preferentially grows $\bm{B}_{\perp}/\!\sqrt{\rho}$ compared to $\bm{u}_{\perp}$, thus  acting like a reflection term for  largest-scale waves and aiding in the generation of turbulence.
%\footnote{The non-WKB driving terms can be related  to wave reflection from gradients of $\vA$ using mass conservation along a flux tube  \citep[see][]{Heinemann1980,Chandran2009,Tenerani2017}}

%Homogenous fields and flows  decay 
%at different rates in the directions parallel and perpendicular to the expansion: $B_{x}/\sqrt{\rho}\propto a^{-1}$, $B_{\perp}/\sqrt{\rho}\propto a^{0}$, 
%$u_{x}\propto a^{0}$, $u_{\perp}\propto a^{-1}$, with $ \rho\propto a^{-2}$  \citep{Grappin1996}. 
%These  scalings apply to motions that are  slow compared to $\dot{a}$; in the opposite 
%limit of small-amplitude, high-frequency Alfv\'en waves  ($\omega_{A}\gg \dot{a}/a$), the wave amplitude grows as $B_{\perp}/B_{x}\propto a^{1/2}$, $u_{\perp}/\vA\propto a^{1/2}$.

 \subsection{General considerations}\label{sub:general.considerations}
 \label{sec:general}
 
 In order to simulate the propagation of Alfv\'enic fluctuations from the transition region to PSP's first perihelion, our simulations would ideally have the following two properties:
 \itemspace
\textbf{(i) Large expansion factor.} As a parcel of solar-wind plasma flows from $r_{\rm i} \simeq 1 R_{\odot}$ to $r_{\rm f} \simeq 35 R_{\odot}$, it expands by a factor exceeding 35 because of super-radial expansion in the corona, leading (at least in the absence of dissipation) to a large increase in normalized fluctuation amplitudes.
\itemspace
%\item[Slow expansion] At $R=35.7R_{\odot}$, the expansion timescale is $U/R\approx8.5\times 10^{4}\,{\rm s}$, meaning all fluctuations with
%frequency $f\gtrsim 1.2\times 10^{-5}\,{\rm s}^{-1}$ are in high-frequency regime with $\omega_{A}>\dot{a}/a$. 
\textbf{(ii) Minimum dissipation. }The damping of inertial-range Alfv\'enic fluctuations is negligible, aside from nonlinear damping processes (turbulence, parametric decay, etc.).\itemspace
Unfortunately, these properties are difficult to achieve: (i) causes the numerical grid to become highly anisotropic, which can cause numerical instabilities, while (ii)  necessitates large numerical grids and careful choice of dissipation properties. Further, probing smaller scales deeper in the inertial range (as commonly done with elongated boxes in turbulence studies; \citealp{Maron2001}) necessitates a slower expansion rate, which makes (ii) even more difficult to achieve because of the long simulations times.
 Our parameter scan and numerical methods are designed to investigate the basic physics of solar-wind turbulence within these limitations.
 
%%%%%%%%%%%%%%%%%%%%%%%%%%%%%%%%%%
\begin{table}%[!htb]
%\captionsetup{size=\footnotesize}
\caption{Properties of the simulations studied in this work ($\beta$ is shown in figure~\ref{fig:PSPtracks}a). \\$L_{\perp}^{i} $ and $L_{\perp}^{f} $ are the initial and final perpendicular box size.} \label{tab:sims}
\setlength\tabcolsep{0pt} % let LaTeX compute intercolumn whitespace
\smallskip 
\begin{tabular*}{\columnwidth}{@{\extracolsep{\fill}}lcccccr}
\toprule
  Name  & Resolution & Code       & $\big(\frac{L_{x}}{L_{\perp}^{i}},\,\frac{L_{\perp}^{f}}{L_{\perp}^{i}}\big)$          & $\frac{\dot{a}}{a}\frac{L_{x}}{\vA}$    & $\frac{\langle B_{\perp0}^{2}\rangle}{B_{x0}^{2}}$             & Spectrum                                  \\
\midrule
SW-GS     &   $480^{3}$ & Snoopy   &  $(4,\,10) $                              &  $2$  &   0.2     &     $k_{\perp}^{-5/3}$, $k_{\parallel}^{-2}$ \\
SW-Gaus&   $480^{3}$ & Snoopy   &  $(4,\,10) $                               &  $2$  &   0.2     &        Gaussian          \\
SW-s1      &   $480^{3}$ & Snoopy   &  $(4,\,10) $                                 &  $2$  &    0.2     &       $k^{-1}$          \\
SW-GS-LR     &   $240^{3}$ & Snoopy   &  $(4,\,10) $                            &  $2$  &   0.2     &        $k_{\perp}^{-5/3}$, $k_{\parallel}^{-2}$ \\
$\beta1$-Gaus-HR    &   $540\times1120^{2}$ & Athena++   &  $(1,\,4) $  &  $0.5$  &    1.0        &   Gaussian   \\
$\beta1$-GS     &   $280\times540^{2}$ & Athena++   &  $(1,\,4) $      &  $0.5$  &   1.0        &       $k_{\perp}^{-5/3}$, $k_{\parallel}^{-2}$ \\
$\beta1$-Gaus     &   $280\times540^{2}$ & Athena++   &  $(1,\,4) $  &  $0.5$  &    1.0        &   Gaussian   \\
$\beta1$-s1     &   $280\times540^{2}$ & Athena++   &  $(1,\,4) $        &  $0.5$  &    1.0        &   $k^{-1}$   \\
$\beta1$-s3     &   $280\times540^{2}$ & Athena++   &  $(1,\,4) $         &  $0.5$  &    1.0        &   $k^{-3}$   \\
$\beta0.1$-GS     &   $280\times540^{2}$ & Athena++   &  $(1,\,4) $      &  $0.5$  &    1.0        &   $k_{\perp}^{-5/3}$, $k_{\parallel}^{-2}$ \\
$\beta0.02$-GS     &   $280\times540^{2}$ & Athena++   &  $(1,\,4) $    &  $0.5$  &   1.0        &    $k_{\perp}^{-5/3}$, $k_{\parallel}^{-2}$ \\
$\beta35$-GS     &   $280\times540^{2}$ & Athena++   &  $(1,\,4) $    &  $0.5$  &    1.0       &   $k_{\perp}^{-5/3}$, $k_{\parallel}^{-2}$ \\
$\beta1$-$\dot{a}0.2$     &   $280\times540^{2}$ & Athena++   &  $(1,\,4) $     &  $0.2$  &    1.0        &   $k_{\perp}^{-5/3}$, $k_{\parallel}^{-2}$ \\
$\beta1$-$\dot{a}0$     &   $280\times540^{2}$ & Athena++   &  $(1,\,1) $        &  $0$  &    1.0       &   $k_{\perp}^{-5/3}$, $k_{\parallel}^{-2}$ \\
\bottomrule
\end{tabular*}
\end{table}
%%%%%%%%%%%%%%%%%%%%%%%%%%%%%%%%%%

\subsection{Numerical methods}\label{sub:numerics}

For the reasons discussed above,  we use two separate numerical methods to solve equations \eqref{eq:mhd.rho}--\eqref{eq:mhd.b}, which have different strengths and provide
complementary results. The first -- a modified version of the pseudospectral code Snoopy \citep{Lesur2007} -- allows for 
 larger expansion factors, making it possible to probe the growth of turbulence and switchbacks from low-amplitude waves. 
 The second -- a modified version of the finite-volume code Athena++ \citep{Stone2008,White2016} -- is better suited for capturing shocks 
 and sharp discontinuities, but develops numerical instabilities for $a\gtrsim4$.
In Snoopy, we solve equations \eqref{eq:mhd.rho}--\eqref{eq:mhd.b} using $\ln\rho$ and $\bm{p}\equiv\rho \bm{u}$ as variables, 
 applying the $\nablaex$ operator by making the $k_{y}$ and $k_{z}$ grids shrink in time. We
 use sixth-order hyper-dissipation to regularize $\bm{p}$, $\bm{B}$, and $\ln\rho$, with 
  separate time-varying parallel and perpendicular diffusion coefficients chosen to ensure that the Reynolds number 
  remains approximately constant as the box expands.
  %\footnote{Specifically, we measure the Reynolds number  at twice the grid scale in one of the simulations (SW-s1) and use this diffusion coefficient for the other simulations.} 
In Athena++, we use  variables  $\bm{B}_{\perp}' = \bm{B}_{\perp}/a$ and $\bm{u}_{\perp}' = \bm{u}_{\perp}/a$, which
  leads to MHD-like equations with a modified magnetic pressure \citep{Shi2019}. 
We use a simple modification of the HLLD Reimann solver of \citet{Mignone2007} with second-order spatial reconstruction, without explicit resistivity or viscosity.

\subsection{Simulation parameters and initial conditions}\label{sub:ics}

Our simulations are listed in Table \ref{tab:sims}. Each starts with randomly phased, ``outward-propagating'' ($z^{+}$) Alfv\'en waves with  $\bm{u}_{\perp}=\bm{B}_{\perp}$ and $u_{x}=0$. The initial normalized fluctuation amplitude  $\langle B_{\perp0}^{2}\rangle/B_{x0}^2=\langle B_{y0}^{2}+B_{z0}^{2}\rangle/B_{x0}^2$ in each simulation varies between $0.2$ and $1$. 
To explore the range of power spectra that might be present in the low corona \citep[see, e.g.,][]{vanBallegooijen2011,Chandran2019}, we consider three types of initial magnetic power spectra $\eperp(k)$ in our simulations: (i) isotropic spectra peaked at large scales, including Gaussian spectra [$\eperp(k)\propto \exp(-k^{2}/k_{0}^{2})$ with $k_{0}=12/L_{\perp}$]  and $\eperp(k)\propto k^{-3}$;  (ii) isotropic spectra with equal power across all scales [$\eperp(k)\propto k^{-1}$]; and  (iii) critically balanced spectra $\eperp(k_{\perp},k_{\|})\propto k_{\perp}^{-10/3}\exp(-k_{\parallel}L_{\perp}^{1/3}/k_{\perp}^{2/3})$, which we denote with the shorthand ``$k_\perp^{-5/3}, k_\parallel^{-2}$'' in Table~\ref{tab:sims}, corresponding to the
 one-dimensional perpendicular and parallel spectra in the critical-balance model  \citep{Goldreich1995,Cho2002}.
We avoid imposing any specific features on the flow or field (e.g., large-scale streams; \citealp{Shi2019}).
%, or directly injecting backwards-propagating Alfv\'en waves that would aid in the development of turbulence \citep{vanBallegooijen2016,Chandran2019}. 
 Two additional  simulation parameters are
 $c_{s}^{2}(t)$, which determines $\beta$ through $\beta=8\pi c_{s}^{2}(t)\langle \rho /B^{2}\rangle$, and the expansion rate
 \begin{equation}\Gamma_{\rm{sim}}\equiv\frac{\dot{a}}{a}\frac{L_{x}}{\vA},
 \label{eq:defGamma}
 \end{equation} 
 which  is the ratio of outer-scale Alfv\'en time $L_{x}/\vA$ to expansion time $(\dot{a}/a)^{-1}$. In our simulations, $\Gamma_{\rm{sim}}$ is constant because $\vA\propto a^{-1}$, while $\dot{a}$ and $L_{x}$ are constant.\footnote{By assuming constant $\dot{a}$, our simulations do not capture the variable expansion rate of the solar wind at small $R$  \citep[see][]{Tenerani2017}. }%This limitation seems unlikely to be important for the physics of interest.}  

 %%%%%%%%%%%%%%%%%%%%%%%%%%%%%%%%%%
\begin{figure*}
\begin{center}
\includegraphics[width=0.95\columnwidth]{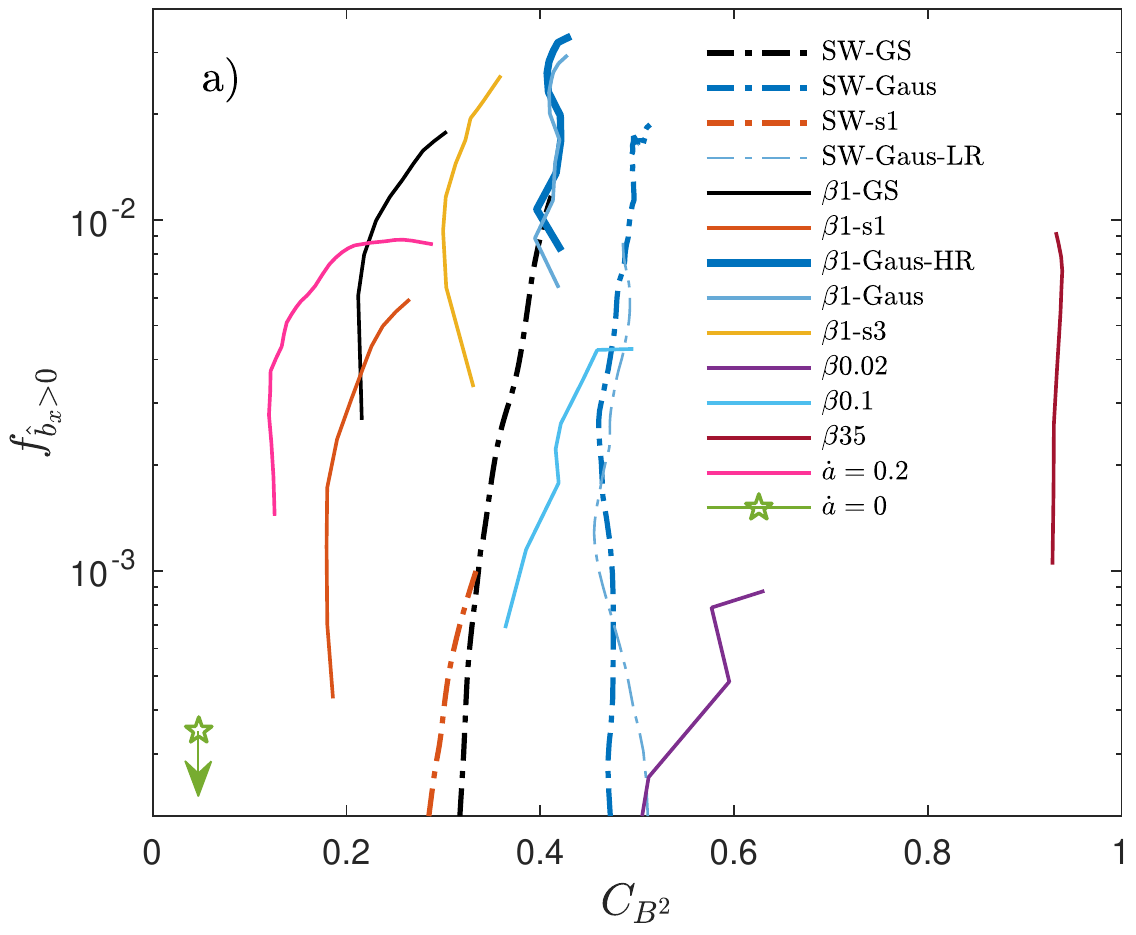}~~~~~~~~\includegraphics[width=0.95\columnwidth]{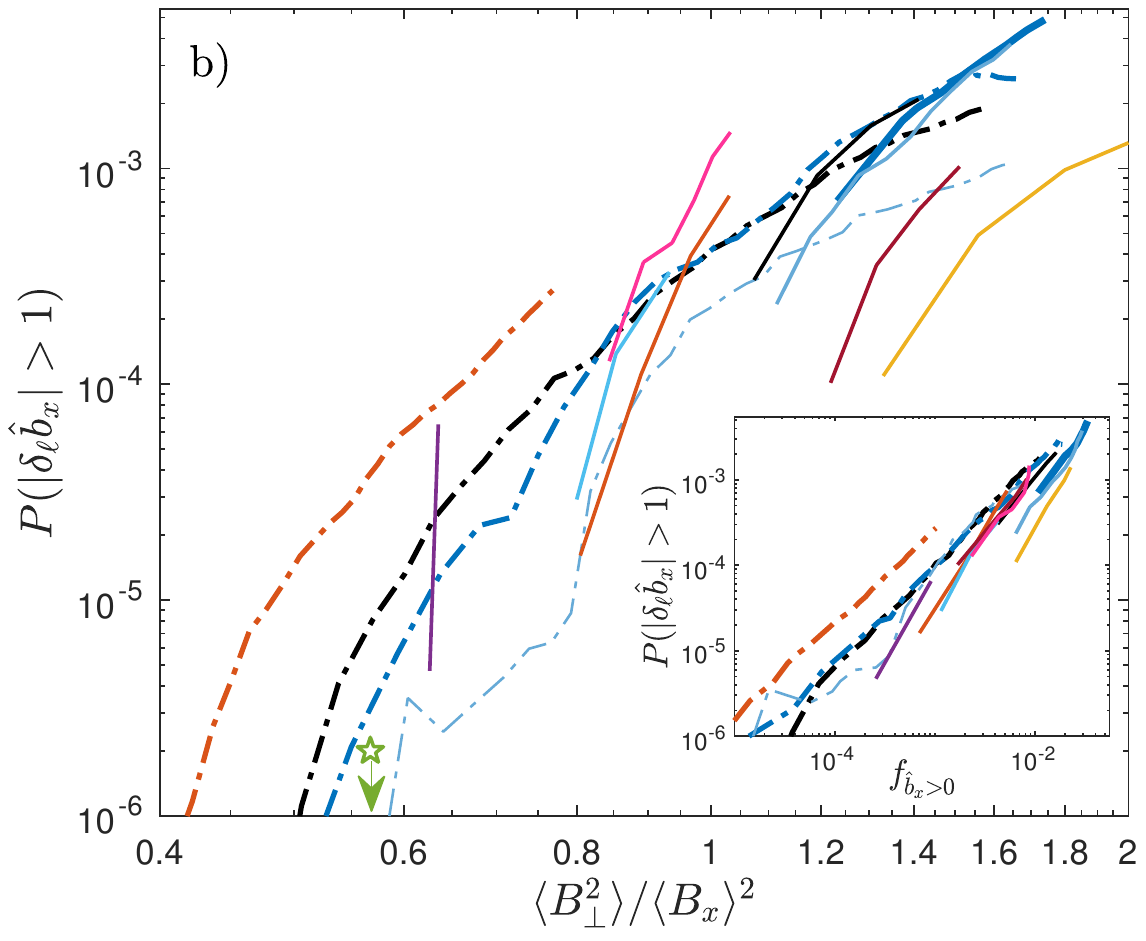}
\caption{Key statistics from all simulations; each line shows the path of a given simulation as labeled 
(lines plot at $a \gtrsim 1.7$ in the Athena++ simulations to avoid the initial transient). Panel a compares the magnetic compressibility $C_{B^{2}}$ to $\sbfrac$, illustrating
how $C_{B^{2}}$ is minimized for $\beta\sim1$ and when the expansion rate is slower.  
Panel b shows how  
%small-scale changes in field-deflection angle, 
abrupt changes in field direction, as measured by $\sbgrad$, 
%grow with  fluctuation amplitude. 
become more common as $\amp$ grows. The 
inset shows the good correlation between the two switchback-fraction measures, $\sbfrac$ and $\sbgrad$. Simulation $\beta1$-$\dot{a}0$ has 
$\sbfrac=\sbgrad=0$ (indicated with a downwards arrow).}
\label{fig:scatter}
\end{center}
\end{figure*}
%%%%%%%%%%%%%%%%%%%%%%%%%%%%%%%%%%

Since most of the fluctuations propagate away from the Sun at speed~$\vA$ in the plasma frame,
we map position~$x$ in the simulation to time~$t$ in the perihelion (zero-radial-velocity) PSP frame by setting
\begin{equation}t =\frac{x}{U+\vA}+\mbox{constant}.\label{eq:tmap}\end{equation}
We then convert from simulation units to physical units by equating $\dot{a}/a$ in the simulations with $U/r$ and setting
$r = 35 R_{\odot}$,  $U=300\,\rm{km}\,\rm{s}^{-1}$, and 
 $\vA = 115\,\rm{km\,s}^{-1}$, values that approximate the conditions during PSP's first perihelion pass \citep{Kasper2019,Bale2019}. The
 parallel box length $L_x$ then corresponds via
 equation~(\ref{eq:defGamma}) to the physical length scale $ 4.8\times10^{6}\Gamma_{\rm{sim}}\,\rm{km}$. An outward-propagating Alfv\'en wave with parallel wavelength~$L_x$ has a frequency
 \begin{equation}
     \frac{U+\vA}{L_x} = 8.7\times10^{-5}\left(\frac{\Gamma_{\rm{sim}}}{0.5}\right)^{-1}\rm{Hz}
     \label{eq:f0}
 \end{equation}
 in the spacecraft frame.  The smallest resolvable parallel wavelength, approximately 4 times the grid scale, corresponds to an Alfv\'en-wave frequency $(N_\parallel/4)\times(U+\vA)/L_x$, where $N_\parallel$ is the number of grid points in the $x$ direction.
 %times larger than
% This length scale, in turn, corresponds to a 
%after converting to a frequency by assuming radial wave propagation at speed $U+\vA$, implies that a simulation's parallel outer scale ($L_x$) corresponds to the solar-wind frequency
%\begin{equation}f_{0}\approx8.7\times10^{-5}\left(\frac{\Gamma_{\rm{sim}}}{0.5}\right)^{-1}\rm{Hz} .\end{equation}
%The smallest resolvable scale, approximately 4 times the grid scale, corresponds to a frequency $f\approx1.2\times10^{-2}({\Gamma_{\rm{sim}}}/{0.5})^{-1}\rm{Hz}$ at the default resolution.

The Snoopy simulations, labeled SW-\#, are designed to optimize property~(i) of Section~\ref{sec:general}.
%to approximate the pathway of solar-wind plasma as it propagates outwards from the sun. 
They start from small-amplitude waves,  $\langle B_{\perp0}^{2}\rangle/B_{x0}^{2}= 0.2$, and expand by a factor of $a_{\rm{f}}/a_{\rm{i}}=10$
to reach large normalized amplitudes. We
prescribe 
non-adiabatic 
temperature evolution $c_{s}^{2}(t)=0.35t^{-1}$ for these runs, causing $\beta$ to increase from $\beta\approx0.2$
to $\beta\approx1$ (figure \ref{fig:PSPtracks}a), in approximate agreement with near-sun  predictions from the model of \citet{Chandran2011} with parameters chosen to match conditions seen by PSP \citep[see][]{Chen2020}.
The amplitude increases broadly as expected, with some dissipation, as shown in figure \ref{fig:PSPtracks}b.
In contrast, the Athena++ simulations (labeled $\beta$\#-\#) are designed to 
explore the basic physics of expanding-box turbulence at a range of $\beta$  values. They are limited
to $a_{\rm{f}}/a_{\rm{i}}\lesssim4$ by numerical instabilities,
which necessitates starting from  larger wave amplitudes, $\langle B_{\perp0}^{2}\rangle/B_{x0}^{2} = 1.0$. 
The fluctuations rapidly (within less than an Alfv\'en time) develop broadband turbulence and near-constant $\B$, resembling that at later times in the Snoopy runs.
The temperature evolution in these runs is adiabatic  ($c_{s}^{2}(t)\propto a^{-4/3}$), so $\beta$ remains nearly constant %over the simulation 
 (figure \ref{fig:PSPtracks}a). 
We choose $\Gamma_{\rm{sim}}=0.5$ for most simulations so that wave growth is mostly in the WKB regime, although it proved necessary to use an elongated box ($L_{x}=4,\,\dot{a}=0.5;\,\Gamma_{\rm{sim}}=2$) in the Snoopy simulations because of the enhanced dissipation over longer simulation times. 
 We also explore a slower expansion rate $\Gamma_{\rm{sim}}=0.2$
($\beta$1-$\dot{a}0.2$) and the same initial conditions with no expansion at all ($\beta$1-$\dot{a}0$).

%%%%%%%%%%%%%%%%%%%%%%%%%%%%%%%%%%%
%\begin{figure}
%\begin{center}
%\includegraphics[width=1.0\columnwidth]{\ffold fig_snoopyt}
%\caption{Time evolution of the SW simulations. The top panel shows the perpendicular turbulence amplitude of magnetic (solid lines) and 
%velocity (dashed lines) fluctuations. The right axis and dotted lines show the normalized cross helicity $\sigma_{c}\equiv \langle \rho \bm{u}\cdot\bm{B}/\rho^{1/2}\rangle/\langle \rho u^{2} + B^{2}\rangle$. The bottom panel shows 
%the magnetic compressibility $C_{B^{2}}$ with solid lines and the left-hand axis, and the relative $B_{x}$ fluctuation with dotted lines on the right-hand axis. 
%The regions of thicker dotted lines show those times for which a fraction of the box larger than $5\times 10^{-4}$ has a magnetic field reversal ($B_{x}<0$).  }
%\label{fig:snoopy.t}
%\end{center}
%\end{figure}
%%%%%%%%%%%%%%%%%%%%%%%%%%%%%%%%%%%

\section{Results}

The evolved state from a selection of runs is shown in figure~\ref{fig:vis}, with a focus on  $\beta$1-Gaus-HR. 
The field-line visualization shows a strong switchback with a complex, three-dimensional structure,  
as well as 
%illustrating  
sharp, 
%angular structures 
angular bends along 
%in
other field lines.
%and the perpendicular fields and flows.
In the spacecraft-like  trace shown below (panel c), 
we see that the switchback (around $t=0$) results in 
a full reversal of the magnetic field to the antisunward direction $\hat{b}_{x}=B_{x}/\B\approx+1$.
%across a scale corresponding to around $\sim\!500\,{\rm s}$. 
The spacecraft remains in the $B_x>0$ region for approximately $1000\rm{s}$.
The jumps at the boundaries of the switchback, with a scale of $\sim\!100\,{\rm s}$, are resolved by only $8$ to $10$ grid cells.
Because of dissipation, this $\sim\!\!100\,\rm{s}$ scale is likely around the simulation's minimum resolvable scale.
%likely around the minimum scale that can realistically 
%be resolved due to grid dissipation. 
Figure~\ref{fig:vis}c also shows 
%We also see 
that $\B$ 
%is kept 
remains fairly constant despite large changes to $\bm{B}$, as ubiquitously observed in solar-wind data \citep{Belcher1971,Barnes1981,Bruno2013}.\footnote{The  small-scale oscillations at
$t\approx 35000\,\rm{s}$ in
figure~\ref{fig:vis}c  result from a numerical instability caused by the anisotropic grid. They appear only for $a\gtrsim 3.5$ and are 
the reason we limit our Athena++ simulations to $a\leq 4$.}
Time traces in the lower panels illustrate several other relevant simulations. SW-Gaus 
%shows similar  behavior to
resembles
$\beta$1-Gaus-HR,
%though on
but with
%somewhat
larger time scales  
%with less sharp  structures 
and smoother variations due to the 
numerical method and lower resolution. In contrast, at high-$\beta$ ($\beta$35-GS), the tendency to maintain constant 
$|\bm{B}|$  is almost eliminated, although there remains significant variation in $B_{x}$. With no expansion ($\beta$1-$\dot{a}0$)
the lack of growth of $\amp$ means there are only small fluctuations in $B_{x}$ and no reversals, while
$\B$ is nearly perfectly constant.

More quantitative measures from all simulations are illustrated 
in figure~\ref{fig:scatter}. We measure ``switchback fraction'' with two methods. First, 
%computing the  field-deflection angle $\cos\alpha\equiv \hat{b}_{x}=B_{x}/\B$ \citep{DudokdeWit2019}, 
we simply count the fraction
of cells with $\hat{b}_{x}>0$, denoting this as $\sbfrac$. Second, to quantify the 
 prevalence of small-scale sharp changes in field direction, we calculate the PDF of $\hat{b}_{x}$ increments, $|\delta_{\ell}\hat{b}_{x}|\equiv|\hat{b}_{x}(\bm{x}+\bm{\ell})-\hat{b}_{x}(\bm{x})|$
across scale $\ell=|\bm{\ell}|=L_{\perp}/68$ ($\sim\!8$ grid cells), and define $\sbgrad$ as the proportion of increments with  $|\delta_{\ell}\hat{b}_{x}|>1$.
%Although imperfect, these provide simple measures of switchback prevalence and (there is not yet clear agreement in the literature;  \citealp{DudokdeWit2019}). 
Two other useful statistics are 
 the ``magnetic compression,'' 
\begin{equation}
C_{B^{2}}\equiv\frac{\delta(\B^{2})}{(\delta\bm{B})^{2}}=\frac{\big\langle\big(\B^{2}-\langle\B^{2}\rangle\big)^{2}\big\rangle^{1/2}}{\big\langle|\bm{B}-\langle\bm{B}\rangle|^{2}\big\rangle},\label{eq:cb2}
\end{equation}
which measures 
%the tendency for the %components of %$\bm{B}$ to align to keep $\B$
the degree of spherical polarization ($|\bm{B}| = \mbox{constant}$),\footnote{Unlike the
commonly used statistic $C_{B}\equiv ({\delta |\bm{B}|})^{2}/{(|\delta \bm{B}|)^{2}}$ \citep{Chen2020}, 
$C_{B^{2}}$ does not decrease with $\amp$ 
for fluctuations $\delta\bm{B}$ oriented perpendicular to~$\bm{B}_0$. Thus, at small $\amp$, $C_{B^{2}}$
still measures correlations between field components, whereas $C_{B}$ becomes a measure of the magnetosonic-mode fraction.
}
and the normalized fluctuation amplitude, which we define as $\langle  B_{\perp}^{2}\rangle/\langle B_{x}\rangle^{2}=
\langle  B_y^{2}+B_z^{2}\rangle/\langle B_{x}\rangle^{2}$. We  choose $\langle B_{x}\rangle$ to normalize the fluctuation amplitudes, as opposed 
to $\langle B_{x}^{2}\rangle$ or $\langle|\bm{B}|\rangle$, because $\langle B_{x}\rangle\propto a^{-2}$
is predetermined by the expansion and $\langle B_{\perp}^{2}\rangle/\langle B_{x}\rangle^{2}$ is not bounded from above \citep{Matteini2018}.

Figure \ref{fig:scatter}a plots $C_{B^{2}}$ and $\sbfrac$ for all simulations
%. We see only 
and shows only modest 
changes to $C_{B^{2}}$ as the fluctuations evolve. %\emph{viz.}, $\B$ is kept remarkably constant, even  with  violent fluctuations in~$\bm{B}$. 
Comparing our
primary set of $\beta\approx 1$ runs with $\beta0.02$, $\beta0.1$, and $\beta35$, we
see -- surprisingly -- that $C_{B^{2}}$ is minimized around $\beta\sim 1$. 
%We speculate that this surprising result may be caused by parametric decay at low $\beta$ and thermal pressure forces at high $\beta$. 
We also observe that $C_{B^{2}}$  increases with
 expansion rate (explaining the larger $C_{B^{2}}$ in Snoopy runs), probably due to non-WKB growth forcing $B_{\perp}>u_{\perp}$ at large $\Gamma_{\rm{sim}}$, 
 and that  $C_{B^{2}}$ increases as the initial spectrum becomes steeper.

\begin{figure}
%\begin{center}
\includegraphics[width=0.98\columnwidth]{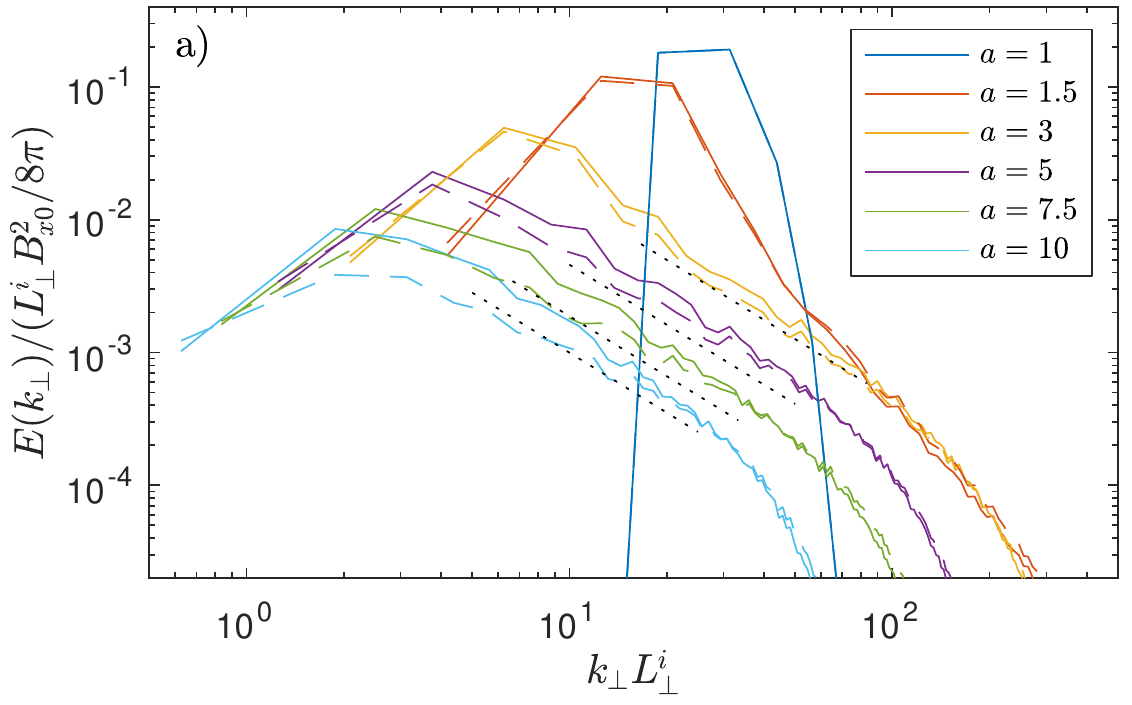}\\\includegraphics[width=1.0\columnwidth]{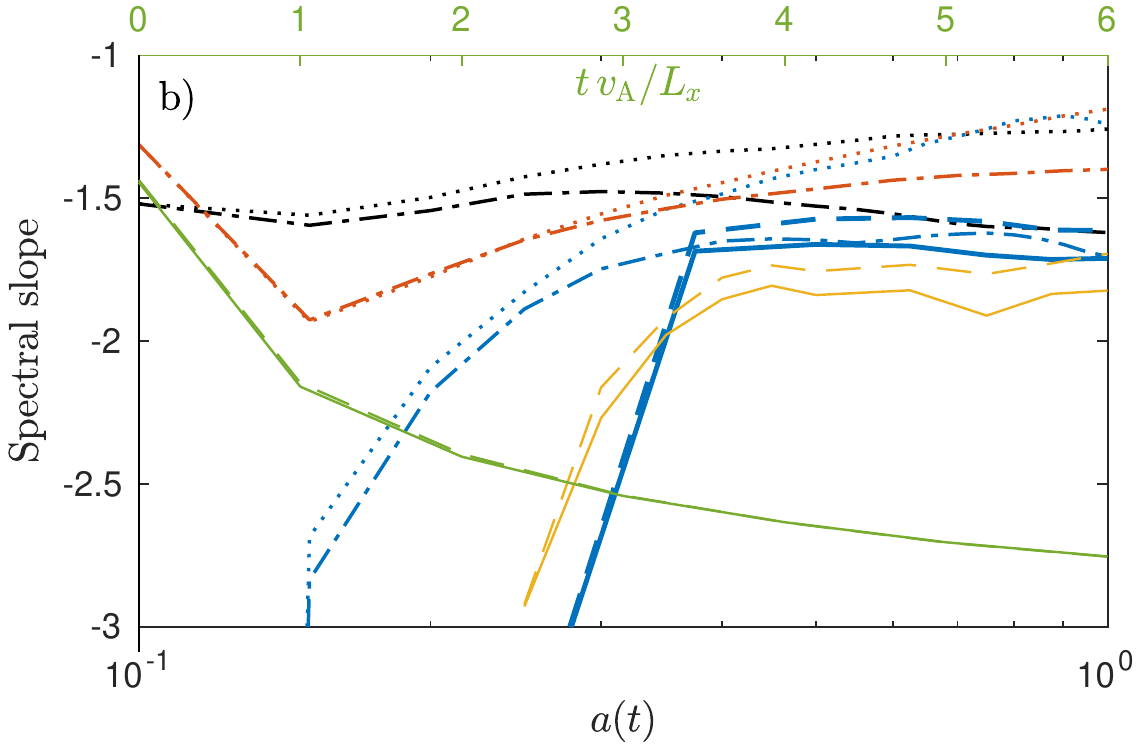}
\caption{Broadband turbulence develops as $\amp$ increases. Panel a shows the
magnetic (solid lines) and kinetic (dashed lines) perpendicular energy spectra at 
%a variety of expansions 
six different times 
%for 
in SW-Gaus. 
The initially  steep (isotropic, Gaussian) spectrum develops a clear power law $\sim k_{\perp}^{-1.5}$ (dotted black lines). Panel b compares the evolution of the spectral slope measured between $k_{\perp}=50/a(t)$ and $k_{\perp}=250/a(t)$ for 
various simulations (line styles as in figure \ref{fig:scatter}). There is a clear evolution towards spectra 
around $k_{\perp}^{-1.5}$, with shallower velocity spectra (dotted lines and dashed lines for Snoopy and Athena++ cases, respectively). 
The green curve with the top time axis shows $\beta1$-$\dot{a}0$, which -- although it exhibits very low magnetic compressibility -- does not evolve towards solar-wind spectra, and does not develop an excess of magnetic energy. 
}
\label{fig:spectra}
%\end{center}
\end{figure}

Figure \ref{fig:scatter}b plots normalized fluctuation amplitude and $\sbgrad$, 
illustrating that the increasing $\langle B_{\perp}^{2}\rangle$  caused by expansion causes the field to become more 
discontinuous. The
good correlation with $\sbfrac$ (inset) shows that similar processes lead to radial-field reversals. 
Comparing $\sbgrad$ between cases $\beta35$-GS and $\beta1$-GS, which have identical initial wave spectra but very different $C_{B^{2}}$, 
 tentatively supports the hypothesis that magnetic-pressure forces, in striving to keep $\B$ constant  \citep{Vasquez1998,Roberts2012},
lead to a discontinuous field with switchbacks.\footnote{$\beta1$-s3, which also has lower $\sbgrad$, has a somewhat steeper spectrum with less overall power at small scales; see figure~\ref{fig:spectra}.} 
We also see that: (i) at low $\beta$ or with shallower initial spectra, the solutions are similarly discontinuous but  the lower $\sbfrac$ and $\sbgrad$ is caused by 
these simulations not reaching such large amplitudes due to increased numerical dissipation  (see also figure \ref{fig:PSPtracks}b); 
(ii) lower resolution simulations have lower switchback fraction, especially in the Snoopy set (SW-Gaus-LR); and (iii)  slower expansion ($\beta1$-$\dot{a}0.2$) causes larger $\sbfrac$ and  $\sbgrad$  for similar amplitudes.

Magnetic-field and velocity spectra are shown in figure \ref{fig:spectra}. Figure~\ref{fig:spectra}a, which pictures SW-Gaus, 
shows how the spectrum flattens to around  $E(k_{\perp})\sim k_{\perp}^{-1.5}$. This is expected and in line with previous results 
because  non-WKB growth from expansion acts like a reflection term for large-scale waves, aiding in the development of turbulence \citep{Grappin1996}.
Figure \ref{fig:spectra}b shows the evolution of the spectral slope from a number of simulations, illustrating the general trend towards 
slopes around $\sim k_{\perp}^{-1.5}$. Although there remain some differences depending on  initial conditions, and a
larger difference between kinetic and magnetic spectra than seen in the solar wind \citep{Chen2020}, %the general agreement is reasonable given 
the spectra in runs with nonzero $\dot{a}$ are quite similar to solar-wind spectra, despite 
the limited 
resolution of the simulations. 
In contrast, 
 simulation $\beta1$-$\dot{a}0$ exhibits continual steepening of the spectrum due to grid dissipation and shows no tendency 
to develop solar-wind-like turbulence (this is also true with $\dot{a}=0$ and large-amplitude $\langle B_{\perp0}^{2}\rangle/B_{x0}^{2}=4$ initial conditions, which have $\sbfrac\neq0$; not shown). 
Finally, we note that the turbulence exhibits scale-dependent anisotropy similar to standard Alfv\'enic turbulence (\citealp{Goldreich1995}; not shown).

%%%%%%%%%%%%%%%%%%%%%%%%%%%%%%%%%%
%%%%%%%%%%%%%%%%%%%%%%%%%%%%%%%%%%

%%%%%%%%%%%%%%%%%%%%%%%%%%%%%%%%%%
\begin{figure}
\begin{center}
\includegraphics[width=1.0\columnwidth]{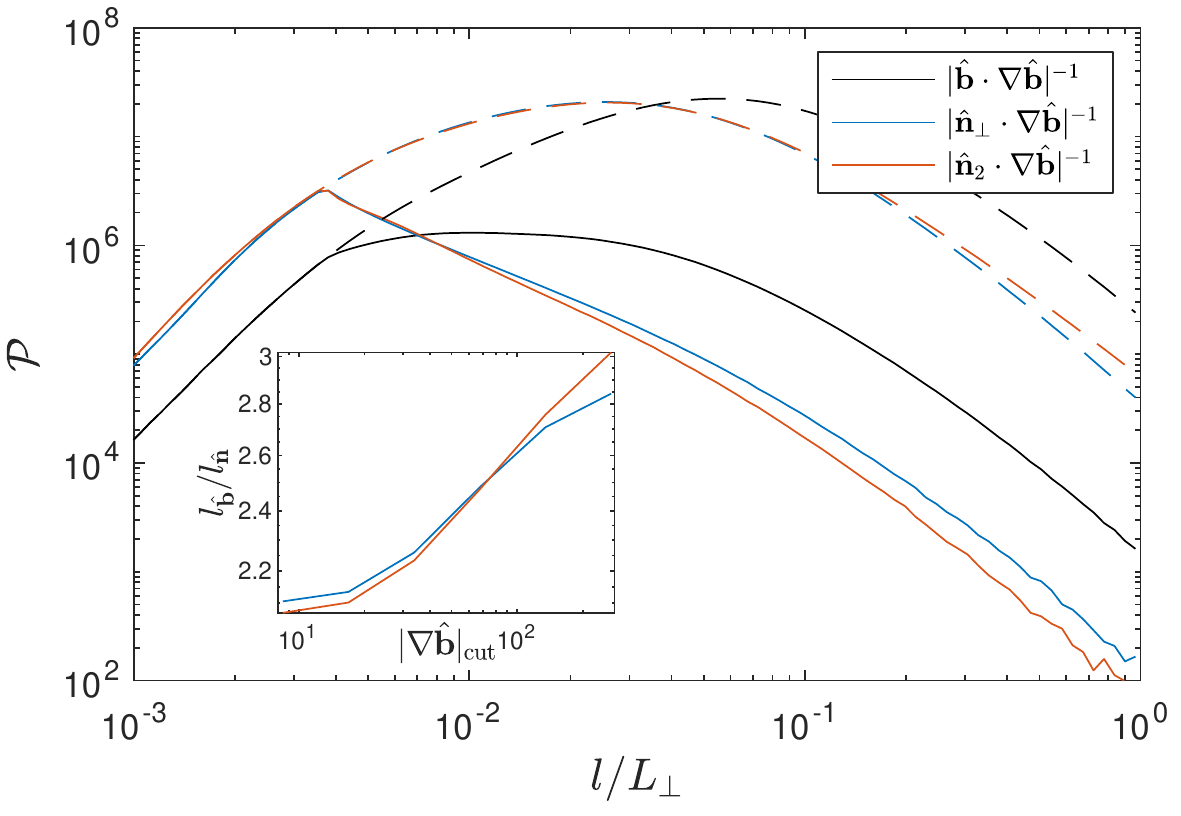}
\caption{The statistical shapes of the most intense, switchback-like structures seen in the highest-resolution simulation $\beta1$-Gaus-HR. 
The main figure shows the PDF of the three lengthscales associated with each structure: field-line parallel, $l_{\hat{\bm{b}}}=|\hat{\bm{b}}\cdot\nabla\hat{\bm{b}}|^{-1}$,
perpendicular in the $y$-$z$ plane, $l_{\hat{\bm{n}}_{\perp}}=|\hat{\bm{n}}_{\perp}\cdot\nabla\hat{\bm{b}}|^{-1}$, and the mutually perpendicular direction, $l_{\hat{\bm{n}}_{2}}=|\hat{\bm{n}}_{2}\cdot\nabla\hat{\bm{b}}|^{-1}$.  Solid lines show sharp structures, those with $|\nabla\hat{\bm{b}}|^{-1}\leq|\nabla \hat{\bm b}|_{\rm cut}^{-1}=4L_{\perp}/N_{\perp}$, while dashed 
lines show all structures with  $|\nabla\hat{\bm{b}}|^{-1}\leq32\times4L_{\perp}/N_{\perp}\approx0.1L_{\perp}$. The PDF of $|\hat{\bm{b}}\cdot\nabla\hat{\bm{b}}|^{-1}$ does not decrease 
so fast with sharpness as $|\nabla\hat{\bm{b}}|^{-1}$, showing that the most intense structures are also more elongated in $\hat{\bm{b}}$, \emph{viz.,} they
are tangential, as opposed to rotational, discontinuities. The inset shows $l_{\hat{\bm{b}}}/l_{\hat{\bm{n}}_{\perp}}=\langle |\hat{\bm{n}}_{\perp}\cdot\nabla\hat{\bm{b}}|/|\hat{\bm{b}}\cdot\nabla\hat{\bm{b}}|\rangle$ (blue line) and $l_{\hat{\bm{b}}}/l_{\hat{\bm{n}}_{2}}=\langle |\hat{\bm{n}}_{2}\cdot\nabla\hat{\bm{b}}|/|\hat{\bm{b}}\cdot\nabla\hat{\bm{b}}|\rangle$ (red line)
as a function of the structure's intensity, again measured by including only those structures with $|\nabla \hat{\bm b}|>|\nabla \hat{\bm b}|_{\rm cut}$.
}
\label{fig:statistics}
\end{center}
\end{figure}
%%%%%%%%%%%%%%%%%%%%%%%%%%%%%%%%%%

Figure \ref{fig:statistics} illustrates the shape of the most  intense structures in $\beta1$-Gaus-HR (other cases are similar). 
We compute $\bm{n}_{\perp} = \hat{\bm{x}}\times \hat{\bm{b}}$ and $\bm{n}_{2} = \hat{\bm{n}}_{\perp}\times \hat{\bm{b}}$, 
and use these to calculate the PDF of structures' length scales along  field lines ($|\hat{\bm{b}}\cdot\nabla\hat{\bm{b}}|^{-1}$) and in the perpendicular directions 
($|\hat{\bm{n}}_{\perp}\cdot\nabla\hat{\bm{b}}|^{-1}$ and $|\hat{\bm{n}}_{2}\cdot\nabla\hat{\bm{b}}|^{-1}$). 
A rotational discontinuity is characterized by the scale of variation along $\hat{\bm{b}}$ being comparable to variation in the other directions, 
while a tangential discontinuity varies more rapidly in the directions perpendicular to $\hat{\bm{b}}$. 
By filtering to include only regions of very large gradient
 as measured by the Frobenius norm ($|\nabla\hat{\bm{b}}|^{-1}\leq|\nabla\hat{\bm{b}|}_{\rm cut}^{-1}=4L_{\perp}/N_{\perp}$; solid lines), 
we see that many of the most intense structures do 
not exhibit sharp changes along $\hat{\bm{b}}$, as evidenced by the lack of a sharp peak in the PDF of $|\hat{\bm{b}}\cdot\nabla\hat{\bm{b}}|^{-1}$ at 
$|\nabla\hat{\bm{b}}|_{\rm cut}^{-1}$ (unlike the PDFs of $|\hat{\bm{n}}\cdot\nabla\hat{\bm{b}}|^{-1}$). This shows that the most intense structures are generally tangential discontinuities,
% as opposed to rotational discontinuities, 
%as also seems to be the case in the field-line visualization of figure \ref{fig:vis}. 
consistent with the simulated fly-by and switchback illustrated in 
figure \ref{fig:vis}a.

\section{Discussion}

The key question that arises from our study is whether 
%the results explain MHD-scale near-Sun turbulence as observed by PSP.  
expansion and spherical polarization generically cause Alfv\'en waves propagating outward from the sun to develop abrupt radial-magnetic-field reversals by the time they reach $r=35 R_{\odot}$, even if the initial wave field is smooth.
%Such understanding 
The answer to this question
is crucial for  assessing what can be 
%learned 
inferred about chromospheric and coronal processes
from PSP 
%results
measurements. Although 
%one
we cannot rule out switchback formation in the low 
solar atmosphere, the simplicity of the in-situ 
formation  hypothesis is compelling, so long as
it is consistent with observations.
%results do 
%it does not explicitly disagree
%with observations. 
%Unfortunately, given the solar wind's  enormous range of scales, minimal dissipation of Alfv\'enic perturbations, and large expansion,  simulations that approach realistic conditions are numerically challenging. 
Our calculations clearly reproduce important basic features,
including sudden 
Alfv\'enic 
radial-field reversals and jets, constant-$\B$ fields, and spectra  around $E(k)\sim k^{-1.5}$. 
%One notable, more subtle, observational characteristic is the presence of long quiet periods between  switchbacks \citep{DudokdeWit2019}\textcolor{red}{HORBURY}, and whether this is reproduced by our simulations 
%remains unclear. While they do
%  produce an  intermittent fluctuation distribution naturally from the wave evolution (compare, e.g., $3\times10^{4}\,\rm{s}\lesssim t\lesssim 10^{5}\,\rm{s}$ with $-8\times10^{4}\,\rm{s}\lesssim t\lesssim 2\times10^{4}\,\rm{s}$ in figure \ref{fig:vis}d), it is 
%also  feasible that quiet periods are produced by global changes, such as low expansion of a particular region or changes in the radial solar-surface field. 
%More quantitatively, 
On the other hand, our simulations % (which reach $\sbfrac\approx2\rightarrow3\%$), 
do not quite reach $\sbfrac\approx6\%$, as  observed in PSP encounter 1 \citep{Bale2019}. 
This discrepancy, however, may simply result from lack of  expansion or scale separation: $\sbfrac$ depends on resolution (figure~\ref{fig:scatter}), particularly for the Snoopy runs, which were designed to approximate the evolution of a solar-wind plasma parcel.  
Similarly, simulation $\beta1$-$\dot{a}0.2$, which  probes smaller scales,\footnote{The larger dissipation in $\beta1$-$\dot{a}0.2$, due to longer integration time, causes $\sbfrac$  to be lower than the similar $\beta1$-GS simulation.} exhibits 
%a 
larger $\sbfrac$ and $\sbgrad$ for similar $\amp$.
More detailed comparison to PSP fluctuation statistics will be left to future work \citep{DudokdeWit2020,Chhiber2020,Horbury2020}.
%Enhanced  intermittency (e.g., extended quiet periods in the solar wind) could also contribute to the discrepancy, since $\sbfrac$  is a highly intermittent statistic, measuring particularly extreme events.

The range of  magnetic compressibility $C_{B^{2}}$ in our simulations broadly matches PSP observations on the
relevant scales. A more intriguing prediction is the dependence of $C_{B^{2}}$ on $\beta$ (figure \ref{fig:scatter}a), which
may be a fundamental property of MHD 
%turbulent evolution
turbulence. While the tendency for magnetic pressure to reduce $\delta\B$ 
has been studied in a variety of previous works \citep[e.g.,][]{Cohen1974,Barnes1981,Vasquez1996,Vasquez1998,Matteini2015}, we believe that our observation
that this tendency is strongest at $\beta\sim1$ is new. The increase in $C_{B^{2}}$  at high $\beta$ is unsurprising:
 thermal pressure forces dominate and interfere with the magnetic-pressure driven 
motions that would act to reduce $\delta\B$ (absent extra kinetic effects; \citealp{Squire2019}).
At low $\beta$, however, the larger $C_{B^{2}}$ is more puzzling; we speculate on two 
possible reasons. First, the theory of \citet{Cohen1974}  for Alfv\'enic discontinuities predicts 
a nonlinear time  that approaches zero at $\beta\sim1$ because  magnetic pressure resonantly 
drives parallel compressions that reduce $\delta\B$. It is, however, unclear how this mechanism operates in a spectrum of oblique waves. Second, 
%and more plausibly, 
predicted parametric-decay growth rates  increase at low
$\beta$  \citep{Sagdeev1969,Goldstein1978,Malara2000,DelZanna2001}, which may disrupt the constant-$\B$  state and  increase $C_{B^{2}}$ \citep[although this remains unclear; see][]{Cohen1974a,Primavera2019}. Parametric decay processes may also 
%source 
aid in the development of turbulence \citep{Shoda2019}.
This physics, along with the role of magnetic pressure in producing discontinuous fields and switchbacks, will be investigated in future work.

%The physics of  switchback formation  and how it relates to $C_{B^{2}}$ also  remains uncertain. Our 
% hypothesis -- that magnetic-pressure forces, by forcing constancy of $\B$ \citep{Vasquez1998}, 
% drive the formation of discontinuities once the amplitude becomes large \citep{Roberts2012,Valentini2019} -- is broadly supported by
% the lower $\sbgrad$ of run $\beta$35 compared to similar amplitude turbulence at lower $\beta$. However, 
% some switchbacks do still develop even though $C_{B^{2}}\approx1$, so there may also be other mechanisms at play. This physics 
% and the low-$\beta$ behavior of $C_{B^{2}}$ will be investigated in future work.

\section{Conclusion}

We show, using expanding-box compressible MHD simulations,
%in the expanding box, 
that a spectrum of low-amplitude Alfv\'en waves 
propagating outwards from the sun naturally develops into turbulence with many similarities to that observed in the recent perihelion passes 
of Parker Solar Probe. Some features that are reproduced by our simulations include  the 
presence of abrupt radial-magnetic-field reversals (switchbacks) associated with 
%jets,
jumps in radial velocity, 
%a reduction in the variation in $\B$ due to correlations between components of $\bm{B}$,
nearly constant magnetic-field strength, and energy spectra with slopes around $k_{\perp}^{-1.5}$.
We present two sets of simulations with complementary numerical methods: the first (SW-\# in table~\ref{tab:sims})
follows a parcel of plasma 
%across a large expansion factor, 
through a factor-of-10 expansion, with waves growing from linear to nonlinear amplitudes; the second ($\beta$\#-\#) explores
the dependence of expansion-driven turbulence on %parameters (initial conditions, $\beta$, and expansion rate). 
initial conditions, $\beta$, and expansion rate.

Our key results can be summarized as follows:\itemspace
(i) Switchbacks can 
form ``in situ'' in the expanding solar wind 
from low-amplitude outward-propagating Afv\'enic fluctuations that grow in normalized amplitude due to radial expansion. This indicates that PSP observations are broadly consistent with  
a natural state 
of large-amplitude imbalanced Alfv\'enic turbulence. The strongest switchbacks are tangential discontinuities. \itemspace
(ii) The magnetic field develops correlations between its components
in order to maintain constant $\B$ \citep{Vasquez1998}. We show that this effect, which has been extensively 
documented in observations, is strongest at $\beta\sim 1$, absent at high $\beta$, and reduced for $\beta\ll1$.\itemspace
(iii) Imbalanced Alfv\'enic turbulence, with energy spectra $\sim\!k_{\perp}^{-1.5}$ and scale-dependent anisotropy, develops from 
homogenous, random, outward-propagating Alfv\'en waves in the compressible expanding box model.
 %The expanding box  provides a useful complement to more complex global models \citep{Shoda2019}.
 \itemspace
We hypothesize that (i) results from (ii) -- i.e., that switchbacks result from the combination of spherical polarization and expansion-induced growth in $\amp$  -- 
because discontinuities are a natural result of large-$\amp$ fluctuations with constant $\B$ \citep{Roberts2012,Valentini2019}. 

To more thoroughly assess the in-situ switchback-formation hypothesis and compare detailed statistical 
measures 
with observations, 
two important numerical goals for future work are  to run to larger expansion factors without excessively fast expansion or high dissipation, and
to reach the steady state where turbulent dissipation 
 balances  expansion-driven wave growth. This will require long simulation times at  high resolution. 
Including mean-azimuthal-field growth (the Parker spiral) 
and/or improved expansion models \citep[e.g.,][]{Tenerani2017} may also be necessary. Theoretically, significant questions remain regarding 
the mechanisms for reducing $\delta\B$ \citep{Cohen1974,Vasquez1998}.

\acknowledgements
We thank Stuart Bale, Chris Chen, Tim Horbury, Justin Kasper, Lorenzo Matteini, Anna Tenerani, and Marco Velli for helpful discussions.
Support for JS and RM was
provided by Rutherford Discovery Fellowship RDF-U001804 and Marsden Fund grant UOO1727, which are managed through the Royal Society Te Ap\=arangi. VC was supported in part by NASA grants NNX17AI18G and 80NSSC19K0829  
and NASA grant NNN06AA01C to the Parker Solar Probe FIELDS Experiment. High-performance computing
resources were provided by the  New Zealand  eScience Infrastructure (NeSI)  under project grant uoo02637 and
the PICSciE-OIT TIGRESS High Performance Computing Center and Visualization Laboratory at Princeton University.  This work was started at the Kavli  Institute  for  Theoretical Physics in Santa Barbara, which is supported by the National Science Foundation under Grant No. NSF PHY-1748958.

%\bibliographystyle{apj}
%\bibliography{fullbib_formatted}

%%%%%%%%%%%%%%%%%%%%%%%%%%%%%%%%%
%%%%%%%%%%%%%%%%%%%%%%%%%%%%%%%%%
%%%%%%%%%%%%%%%%%%%%%%%%%%%%%%%%%
%%%%%%%%%%%%%%%%%%%%%%%%%%%%%%%%%

\end{document}